\documentstyle[twocolumn,aps,epsfig]{revtex}
\begin{document}
\draft
\title{$\Theta$  vacua states in heavy ion collisions in presence
of dissipation and noise}
\author{\bf A. K. Chaudhuri\cite{byline}}
\address{ Variable Energy Cyclotron Centre\\
1/AF,Bidhan Nagar, Calcutta - 700 064\\}
\noindent
\maketitle

\begin{abstract}

We have studied possible formation of $\Theta$ vacua states in 
heavy ion collisions.
Random phases of the chiral fields were evolved in a
finite  temperature  potential,  incorporating  the  breaking  of
$U_A(1)$ symmetry. Initial random  phases  very  quickly  settle
into oscillation around the values dictated by the potential. The
simulation study indicate that an initial $\Theta$=0 state do not
evolve  into  a  $\Theta$  $\neq$  0  state.  However, an initial
$\Theta$ $\neq$ 0 state, if formed in heavy  ion  collision,  can
survive,  as  a  coherent  superposition  of  a  number  of modes.
\end{abstract}


\section{Introduction}

QCD  in  the  chiral limit ($m_u=m_d=m_s=0$) possesses a $U_A(1)$
symmetry. Spontaneous breaking of the symmetry requires a neutral
pseudoscalar  Goldstone  boson  with  mass  less  than  $\sqrt{3}
m_\pi$,  in  addition  to  pion itself. However no such Goldstone
boson is  seen  in  nature.  The  problem  was  resolved  by  the
discovery  of  non perturbative  effects  that violates the extra
$U_A(1)$ symmetry. 't Hooft showed that because of the  instanton
solutions of the Yang-Mills theory, the axial anomaly has nonzero
physical  effects,  and there is not really a U(1) \cite{thooft}.
Term containing the  so  called  vacuum  angle  ($\Theta$)  which
breaks  $P$  and  $CP$  symmetry  can be added to the Lagrangian. QCD
requires very small $\Theta  \sim  10^{-9}$,  which  explain  the
apparent  $P$  and  $CP$  symmetries  in strong interaction Dynamical
breaking of the $U_A(1)$ symmetry is also  obtained  in  large  $N$
(color) limit of the SU(N) gauge theory \cite{wi79,wi80,ve80}. In
this  approach  the  dominant fluctuations are not semi classical
but of quantum nature.

Recently  Kharzeev,  Pisarski  and Tytgat \cite{kh98} and Kharzeev and
Pisarski \cite{kh99} argued
That, in heavy ion collisions, a non trivial $\Theta$ vacua state
may be created. In the limit of large number of colors, the axial
$U_A(1)$ symmetry of massless  quarks  may  be  restored  at  the
deconfining  phase  transition.  As  the  system  rolls  back  to
confining phase, it may settles into a metastable state with non-
trivial $\Theta$.  The  idea  is  similar  to  the  formation  of
disoriented   chiral   condensate   (DCC)   \cite{dcc}.  In  DCC,
space-time region is created, where the chiral condensate  points
in  a  direction,  different  from  the  ground  state direction.
Similarly, in a $\Theta$ vacua  states,  space-time  region  with
non-trivial $\Theta$ is created. $\Theta$ vacua state being P and CP
odd,  such a state, if produced  in heavy ion
collisions,  may have some interesting signals, e.g. decay of $\eta$ into
two pions, which is strongly forbidden in our world.

After  the  suggestion of Kharzeev and co-workers \cite{kh98,kh99}, several
authors have looked into various aspects of nontrivial $\Theta$
vacua  state,  that  may  be  formed  in  heavy  ion  collisions.
\cite{ah00,vo00,bu00a,bu00b}.  Buckley  {\em et  al.} \cite{bu00a,bu00b}
numerically simulated the formation  of  $\Theta$  vacua  states,
using   the  effective  Lagrangian  of  Halperin  and  Zhitnitsky
\cite{ha98}. They assumed quenchlike scenario,  rapid  expansion
of  the  fireball  leave  behind  an effectively zero temperature
region, interior of which  is  isolated  from  the  true  vacuum.
Starting  from an initial nonequilibrium state, they studied the
evolution of  phases  of  the  chiral  field,  in  a  dissipative
environment.  They  saw  formation  of  nonzero  $\Theta$ vacuum
within a time scale of $10^{-23}$ sec. They choose to ignore  the
fluctuation-dissipation theorem, which require the dissipation to
be  associated  with  fluctuations.  Recently we have performed a
simulation  study  for  $\Theta$ vacua  formation  in  heavy  ion
collisions  \cite{ch01}.  The  dissipative term was omitted. In a
quench like scenario, random chiral phases were evolved in a zero
temperature potential. It was shown that, for non-zero  $\Theta$,
randomly  distributed  initial  phases oscillate about the values
dictated by the potential. But, for the physically important case,
the initial $\Theta$=0 state do not evolve into a $\Theta$ $\neq$
0 state.

Quench scenario is not expected to occur in heavy ion collisions.
Unlike  in a quench scenario, where the system is instantaneously
brought  to  zero  temperature,  in  heavy  ion  collisions,  the
fireball  requires a finite time to cool. In the present paper, we
have extended  our  study  \cite{ch01}  to  include   the temperature
dependence of the potential. We have  also added a
dissipative term. Even in  a  quenchlike  scenario,  where  the
chiral  phases  are  evolving,  in  essentially a zero temperature
potential,  a dissipative term is  important.  It
can  mimic  the  effects  like  interaction  between  the quarks,
emission    of    pions    etc.    To    be    consistent    with
fluctuation-dissipation  theorem,  we  have   also   included   a
fluctuation term (which was omitted by  Buckley  {\em et  al.} \cite{bu00a,bu00b}). 
We represent it by  a white noise source.
In a realistic situation, if within a certain region, $\Theta$ vacua
state is formed, that region will be in contact with some environment
(mostly pions etc.). Use of white noise source as the fluctuation term
can take into account those interactions. In Ref.\cite{ch01}, we have
considered a single event, which may be classified as a average type
of event.
 $\Theta$ vacua state being
a rare event, it is not expected to be produced in each and every event.
In the present work, we have classified the events to single out the event 
where $\Theta$ vacua state is most probable. As will be shown below,
even in the most favored event, a initial $\Theta=0$ state, do not evolve into a $\Theta \neq 0$
state.

The  paper  is organized as follow: in section 1, we describe the
model. The results are discussed in section 2.  The  summary  and
conclusions are drawn in section 4.

\section{The model}

As  in  \cite{ch01},  we  have  used  the  effective  Lagrangian,
developed by Witten \cite{wi80},  to  study  the  $\Theta$  vacua
state  in  heavy  ion  collisions. The effective non-linear sigma
model Lagrangian is \cite{wi80},


\begin{eqnarray} \label{1}
{\cal{L}}=f^2_\pi   (\frac{1}{2}   tr   (\partial_\mu   U^\dagger
\partial_\mu U) + tr(M(U+U^\dagger)) \nonumber  \\
- a (tr \ln U - \Theta)^2)
\end{eqnarray}

\noindent where U is a $3\times 3$ unitary matrix with expansion,
$U=U_0  (1  +  i \sum t^a \pi^a /f_\pi + O (\pi^2))$, $U_0$ being
the vacuum expectation value of U, $t^a$ are  the  generators  of
U(3),  ($Tr  t^a  t^b  =\delta  ^{ab})$ and $\pi^a$ are the nonet of
Goldstone boson fields. $M$ is the  quark  mass  matrix,  which  is
positive,  real  and  diagonal. We denote the diagonal entries as
$\mu_i^2$. They are the Goldstone boson squared  masses,  if  the
anomaly  term  $a$  (a/N in ref.\cite{wi80}) were absent. Because
$M$  is  diagonal  $U$  can  be  assumed   to   be   diagonal   ,
$U_{ij}=e^{i\phi_i} \delta_{ij}$.

In terms of $\phi_i$'s, the potential is,

\begin{equation}
V(\phi_i)=f^2_\pi  (-\sum  \mu^2_i  \cos \phi_i +a/2(\sum \phi_i -
\Theta)^2)  \label{2}
\end{equation}

It  may  be noted that as $\sum \phi_i$ arose from $tr \ln U$, it
is defined modulo $2\pi$. Vacuum expectation values of the angles
$\phi_i$'s can be obtained from the minimization condition,

\begin{equation}
\mu^2_i \sin \phi_i = a (\Theta - \sum \phi_j)
\label{3}
\end{equation}

Witten  \cite{wi80}  had  discussed  in  detail, the solutions of
these coupled non-linear equations. If the equation has only  one
solution then physics will be analytic as a function of $\Theta$.
The  solution  vary  periodically  in  $\Theta$  with periodicity
$2\pi$ \cite{wi80}. Also, this solution must  be  CP  conserving,
whenever  CP  is a symmetry of the equation \cite{wi80}. However,
it may happen that, Eq.\ref{3} has more than one  solution.  Then
the  solutions  are not CP conserving, rather a CP transformation
exchanges them.

At finite temperature, the potential (Eq.\ref{2}) is modified due to
temperature dependence of the anomaly term $a$ and  also 
due to temperature dependence of
the Goldstone
Boson squared masses $\mu^2_i$.
In the  mean  field  type  of  theory,  temperature
dependence of $a$ and $\mu^2_i$ can be written as \cite{kh98,kh99},

\begin{eqnarray}
a(T)=&& a_0 (1-T/T_d)\\
\mu^2_i(T)=&& \mu^2_{i0} \frac{1}{\sqrt{1-T/T_d}},
\end{eqnarray}

\noindent  where  $T_d$  is  the  transition  or  the  decoupling
temperature.  The  subscript  $0$  denote  the  values  at   zero
temperature.    In    the    present    work    we    have   used
$\mu^2_{u0}=(114MeV)^2$,                  $\mu^2_{d0}=(161MeV)^2$
,$\mu^2_{s0}=(687)^2$ and $a_0=(492MeV)^2$ \cite{kh98,kh99}. With
these   parameters,   the   mass   matrix  in  (\ref{2})  can  be
diagonalised to obtain $m_\pi^0 \sim$ 139 MeV, $m_\eta \sim $ 501
MeV  and  $m_{\eta^\prime}  \sim  $  983  MeV,  close  to   their
experimental values.

Existence  of  metastable  states  ($\Theta$ vacua states) can be
argued as follows: the vacuum expectation values  depend  on  the
ratio  $a/\mu^2_i$.  The  ratio  $a/\mu^2_i \sim (T_d - T)^{3/2}$
decreases with temperature. Then as the system rolls  toward  the
chiral  symmetry  breaking,  $a/\mu^2_i$  may  be small enough to
support metastable states. In \cite{kh98}, it was shown that when
$a/\mu^2_1 < .2467$ there is  a  metastable  solution,  which  is
unstable    in   $\pi^0$   direction   unless   $a   <   a_{cr}$,
$a_{cr}/\mu^2_1\sim .2403$.

Appropriate  coordinates  for heavy ion scattering are the proper
time  ($\tau$)  and  rapidity  (Y).  Assuming  boost  invariance,
equation of motion for the phases $\phi_i$'s can be written as,


\begin{eqnarray} \label{4}
\frac{\partial^2    \phi_i}{\partial    \tau^2}   +(\frac{1}{\tau} +\eta)
\frac{\partial    \phi_i}{\partial    \tau}     -\frac{\partial^2
\phi_i}{\partial  x^2}  - \frac{\partial^2 \phi_i}{\partial y^2}+
\sum \mu^2_i sin(\phi_i)  \nonumber \\
- a(\sum \phi_i - \Theta)) = \zeta(\tau,x,y)
\end{eqnarray}

It  is  interesting  to  note  that  in this coordinate system, a
dissipative term which decreases with (proper) time is effective.
In addition, we have included the dissipative term with  friction
coefficient ($\eta$). $\eta$ take into account the interaction between
the  quarks.   Expansion of the system, emission of pions etc. also
contribute to $\eta$. The exact form or value of  $\eta$  is  not
known.  It  is expected to be of the order of $\Lambda_{QCD}$=200
MeV. Temperature dependence may be assumed to be same as that
of quarks and gluons.
$\eta$ for quarks and gluons generally shows  a
$T^3$  type of temperature dependence \cite{th91}. We then assume 
the following form for $\eta$,

\begin{equation}
\eta(T) = \eta_0 (T/T_d)^3
\end{equation}

\noindent  with  $\eta_0$=200  MeV. In Eq.\ref{4}, $\zeta$ is the
noise term. We  assume it to  be  Gaussian  distributed,  with
zero     average     and     correlation     as    demanded    by
fluctuation-dissipation theorem,

\begin{eqnarray}
<\zeta_a(\tau,x,y)> =&& 0\\
<\zeta_a( \tau, x, y) \zeta_b(\tau^\prime , x^\prime , y^\prime)>
=  &&  2  T  \eta \frac{1}{\tau} \delta ( \tau -\tau^\prime ) \nonumber \\
&& \delta( x-x^\prime ) \delta ( y - y^\prime ) \delta_{ab}
\end{eqnarray}

\noindent  where  a,b corresponds to phases $\phi_u$, $\phi_d$ or
$\phi_s$.

Solving  Eq.\ref{4}  require  initial phases ($\phi_i$) and their time derivatives 
($\dot{\phi_i}$) at the initial
time ($\tau_i$). We choose $\tau_i$ =1 fm. 
$\phi_i$ and  $\dot{\phi_i}$  were  chosen
according to following prescription,

\begin{eqnarray}
\phi_i = &R_N/(1 + exp((r-r_{cr})/\Gamma) \\
\dot{\phi_i} = & R_N/(1 +exp((r-r_{cr})/\Gamma)
\end{eqnarray}

\noindent  where  $R_N$  is  a  random number within the interval
$[-2\pi/16,2\pi/16]$. We use $r_{cr}$=20 fm and $\Gamma$=.5 fm.

\section{Results}

The  set  of  partial equations (\ref{4}) were solved on a $64^2$
lattice, with lattice spacing of $a=1 fm$ and  time  interval  of
$a/10$  fm. We also use periodic boundary condition. 
 The cooling and expansion of  the
system  was  taken  through  the cooling law $T(\tau)=T_d \tau_i/
\tau$, decoupling or transition temperature being $T_d$ =160 MeV.
The cooling is rather fast and may not be suitable for  heavy ion collisions.
Still we choose to use it, to be as close as possible to a quench like scenario,
which is ideal for  $\Theta$ vacua state formation.
The  phases  were  evolved    for
arbitrarily  long  time of  20 fm. We have performed simulation for
1000 events. The  idea  is  to  see  whether in any of these events,
initial  random  chiral  phases  can  evolve  into  a  nontrivial
$\Theta$ vacua state.  We  have  considered  two  situation,  (i)
$\Theta$=0 and (ii)$\Theta$=$4\pi/16$, within the radius $r_{cr}$
and  zero  beyond $r_{cr}$. Physically, initial $\Theta$=0 is the
most interesting  case.  In  this  case,  emergence  of  non-zero
$\phi$'s  will be the signature of non-trivial $\Theta$ vacua state formation. It
will  show  that  during  the   roll   down   period,   initially
uncorrelated  chiral  phases get oriented in a certain direction.
The 2nd situation, initial  $\Theta$=4  $\pi$/16,  on  the  other
hand,  will  explore  the  possibility  of  sustaining  a initial
$\Theta$ vacua state in a heavy ion collision.

\subsection{Classification of Events}

$\Theta$ vacua state, being a non-trivial state, should be a rare
event.  We do not expect it to be formed in each and every event.
In the ensemble of 1000 events, $\Theta$ vacua state may be formed
in   a few events only. It is then important to classify the events
such that the rare events, where $\Theta$ vacua state is formed,
  can be distinguished. For  initial  $\Theta$=0,
emergence  of  non-zero  $\phi_i$  is    the signature of a nontrivial 
$\Theta$   state. Then  the  space-time  integrated
value  of  the  phases  can  be  used  to  classify  the  events.
Accordingly,  in  each  event  we  calculate  the  classification
parameter,

\begin{equation}  G=\int \tau d\tau dx dy \mid \sum_i \phi_i \mid
\end{equation}

The event for which G is maximum, will be most probable event for
forming  a  $\Theta$ vacua  state.  In  fig.1, we have shown the
distribution of G for  the  1000  events,  for  the  two  sets  of
calculations,    (a)   initial   $\Theta$=0   and   (b)   initial
$\Theta$=4$\pi$/16. Very interesting behavior is obtained.  While
for  $\Theta$=0,  the  distribution  falls  with $G$, it is sharply
peaked  for  $\Theta$=4  $\pi$/16.  We also note that for initial
$\Theta$=0, $G$ ranges between [0.02-0.75]. 
It is apparent that the while the evolution depend on the initial
conditions, in this case, the phases do not evolve
to large values.   Very small value of $G$, indicate that  initial $\Theta$=0 state
 do not 
evolve into a $\Theta \neq$ 0 state.
 In  contrast,  for
initial $\Theta$=4$\pi$/16, $G$ ranges between [154.1-154.9].
The phases grow to large values in this case.
Also very small difference between the minimum and maximum values
of  G  indicate  that  in this case, evolution of phases is largely 
independent of initial conditions. As will be discussed below, this is
an indication that if formed,  potential can support  a $\Theta$ vacua state.

\subsection{Evolution of spatially integrated phases}

In this section we study the temporal evolution of the spatially
averaged chiral phases. For each trajectory,
spatially   averaged  chiral phases,

\begin{equation}
<\phi_i> =\frac{\int \phi_i(x,y) dx dy}{\int dx dy}, i=u,d,s
\end{equation}

\noindent were calculated.
We also calculate the ensemble average of the spatially averaged phases,

\begin{equation}
<\phi_i>_{en} = \frac{1}{N} \sum_{j=1}^N <\phi_i>_j, i=u,d,s
\end{equation}

In Fig.2, for $\Theta=0$, ensemble average of the spatially averaged phases
$<\phi_i>_{en}$,
along with its fluctuations for the 1000 events are
shown (the black circles with error bars). 
All three phases, $\phi_u$, $\phi_d$ and
$\phi_s$ show qualitatively  similar  behavior.  For  $\Theta$=0,
zero      temperature      potential     is     minimized     for
$\phi_u=\phi_d=\phi_s$=0. Even though  fluctuations are large, ensemble averaged  values  remain  zero throughout  the  evolution.  
The result clearly shows that on  the
average,  initial  $\Theta$=0  state  do not evolve into a $\Theta
\neq$ 0 state. 
 In Fig.2, solid lines shows the evolution of the spatially
averaged  phases ($<\phi_i>$)
in  the  most  favored  event.
 For initial $\Theta$=0, non-zero $\phi_i$'s
will be  indication  of  a  $\Theta$ vacua  state.  It is then expected
that if a $\Theta$ vacua state is formed, $<\phi_i>$'s will remain nonzero for a substantial time. We do not
find such behavior. On the contrary,
spatially  averaged phases ($<\phi_u>$,$<\phi_d>$ and $<\phi_s>$)
oscillate about zero, their vacuum expectation value. Amplitude of the
oscillation decreases with time (due to friction). Even in the most favored
event, initial $\Theta = 0$ state do not evolve into a nontrivial $\Theta$ state.

 In  Fig.3,  same  results  are shown for initial $\Theta$=4 $\pi$
/16. The ensemble  average of spatially averaged phases $<\phi_i>_{en}$,
 do  not  show large
fluctuations. Very narrow range of $G$ already indicated this. As a result,
 $<\phi_i>$ in the most
favored event (the solid lines) closely follow $<\phi_i>_{en}$.
For $\Theta$=4$\pi$/16, the minimization condition
gives, $\phi_u$=0.502, $\phi_d$=0.243 and  $\phi_s$=0.013,  which
are shown by the straight lines in Fig.3, The spatially averaged 
phases are found to
oscillate  about these values throughout the evolution. As will be
discussed  below,  continued  oscillation  of the phases indicate
that in heavy ion collisions, an initial $\Theta$ vacua state  is
sustained as a coherent superposition of a number of modes.

\subsection{Correlation}

We define a correlation function,

\begin{equation}
C(r)=  \frac{\sum_{i,j}  \roarrow{  \phi_i}  .  \roarrow{\phi_j}}
{\sum_{i,j}
|\roarrow{\phi_i}| |\roarrow{\phi_j}|}
\end{equation}

\noindent  such  that the distance between the lattice points $i$
and  $j$  is  $r$.  $C(r)$  specifies  how  the  three  component
$\roarrow{\phi}=(\phi_u,\phi_d,\phi_s)$  at two lattice points are
correlated. In a nontrivial $\Theta$ vacua state, phases will  be
correlated   while   in  a  trivial  $\Theta$ vacua  state  (i.e.
$\Theta$=0) phases will remain uncorrelated. In  Fig.4,  we  have
shown  the  evolution  of the correlation function for $\Theta$=0
and $\Theta$=$4\pi/16$ at different times, $\tau$=1,3,5,7 and  9  fm.
The  results  are  shown for the most favored event. Initially at
$\tau$=1 fm, for both the cases, correlation length  is  about  1
fm,   the   lattice   spacing.   Initial  phases  were  random  ,
uncorrelated. The correlation length increases  at  later  times.
But  for  $\Theta$=0,  the  increase  is  marginal.  Thus initial
uncorrelated phases do not  develop  correlation,  confirming  our
earlier  results  that  $\Theta$=0  state  do not  evolve  into  a
non-trivial $\Theta$ state. However for finite initial  $\Theta$,
correlation  length  increases  very  rapidly,  and assumes quite
large  values.  Physically,   for   finite   $\Theta$,   all   the
$\roarrow{\phi}$'s  try  to  align  themselves  in some direction
(i.e. in the  $\Theta$  direction),  thereby  giving  very  large
correlation  length,  even  when a large distance separates them.
This figure clearly demonstrated the possibility  of  parity  odd
bubbles  formation  in  heavy  ion  collisions.  Initially random
phases evolve such  that  they  points  in  the  same  direction,
forming a large P, CP odd bubble.

\subsection{Momentum space distribution of phases}

Much  insight  to  the  process can be gained from the momentum
space distribution of the phases. At each alternate time step, we
apply a fast Fourier transform to the spatial data.  The  Fourier
transformed  data  are  then integrated over the angles to obtain
momentum distribution,

\begin{equation}
\phi_i(k)=\frac{1}{2\pi}  \int^{2\pi}_0 \phi_i(k,\theta) d\theta,
i=u,d,s
\end{equation}

Each  mode  was  averaged over some narrow bin.  For initial $\Theta =0$,
evolution of ensemble averaged $\phi_i(k)$'s
at   k=6.2   MeV  (which  we  call  zero  mode)  along  with  its
fluctuations are shown (dots with error bars) in Fig.5. Ensemble
averaged $\phi_i(k)$'s remain zero through out the evolution. They
do not grow.
The behavior is in
agreement  with  the  results  obtained  for  ensemble average of
spatially  averaged
$\phi_s$'s. On the average,    initial $\Theta$=0 state do not
evolve into a $\Theta \neq 0$ state.
In Fig.5, evolution  of  the zero  mode (solid line) and higher modes k=18.5 and 30.9 MeV (dash and dash-dashed line respectively), in  the most favored event are also shown. 
The
behavior of the modes in the most favored event is different. The
zero mode as well as higher modes oscillates about  the  ensemble
averaged  zero value. Higher modes are largely suppressed compare
to zero mode. The state is  essential  a  zero  mode  state.  The
results again confirms   that, even in the most favored event, a
trivial  $\Theta$  state do not
evolve into a non-trivial $\Theta$  state. 

In Fig.6, same results are
shown for initial $\Theta$=4$\pi$/16. In contrast  to  $\Theta$=0
case,  
here the ensemble averaged zero
mode quickly reaches some finite value, though it also shows  large
fluctuations.  The  modes in the most favored event show similar
oscillatory behavior,
oscillating about the  ensemble  averaged  value.  And  as  before,
higher modes are suppressed. However, suppression is not enough to label
the state as a zero mode state. 
It  is better to describe the state as a coherent  superposition  of
different  modes. The results are in agreement with the evolution
of  spatially  averaged  $\phi_i$'s.  We  found  that   spatially
averaged  $\phi_i$'s execute oscillation about the value dictated
by the potential. Present simulation  results  are  qualitatively
different   from   the   simulation  results  of  Buckley  et  al
\cite{bu00a,bu00b}. They found  that  for  finite  $\Theta$, after a few oscillations,
 the zero mode  quickly reaches the value dictated by the potential.
Also higher modes are largely suppressed.
The difference is essentially due to the strong dissipative environment
(with out fluctuations),
 in which  the phases were evolved.
In  the  present  case,
friction  assumes  low value at late times, resulting in the
continued oscillation of the phases, even at late times.

Present simulations indicate that in heavy ion collision, initial
random  phases  do  not  evolve into a non-trivial $\Theta$ vacua
state. However, an initial finite  $\Theta$  state,  survive  the
evolution,  as  a  coherent superposition of different modes. The
result is encouraging, as it establishes that if a $\Theta$ vacua
state is created in a heavy ion collision, it will  survive.  This
leaves open the possibility of creating a $\Theta$ vacua state in
heavy  ion  collisions.  Though our study indicate that initial random
phases do not evolve into a non-trivial $\Theta$ vacua state,  it
is not definitive. We have considered 1000 events only. Sample
size may be too small to detect a $\Theta$ vacua state, which,  as
noted  earlier  is  a  rare  state. Also our event classification
scheme may not be the best. It may be possible to devise a better
criterion to distinguish between different events to detect a  $\Theta$
vacua state.

What  will  be  the  signature of such a state. As such detecting
$\Theta$ states are difficult. Kharzeev and Pisarski  \cite{kh99}
devised  some  observables  to  detect  $\Theta$ vacua  state. and
estimated the P-odd observables are of the order of $10^{-3}$,  a
small  effect.  Also  as discussed by Voloshin \cite{vo00} the so
called signal of $\Theta$ states may be faked  by  "conventional"
effects   such   as   anisotropic   flow   etc.   Buckley   et  al
\cite{bu00a,bu00b} suggested the  decay  of  $\eta  (\eta^\prime)
\rightarrow \pi^+ \pi^-$. It is strongly forbidden in our world as C
and P is violated. Thus in a P and CP odd bubble, the decay is a
distinct  possibility. One may look for such decays, which are strongly
forbidden in our world, but have definite probability in a $\Theta$ vacua
state.
 Whatever  be  the  signal of the $\Theta$
vacua states, with a large number of modes  contributing,  signal
will   be   broadened,   effectively   diluting   the   detection
probability.

\section{Summary and conclusions}

To  summarize,  we  have  studied the formation of $\Theta$ vacua
state in a heavy ion collision, in presence  of  dissipation  and
noise.  Assuming  boost-invariance,  equation  of  motion  of the
chiral phases were solved  on  a  $64^2$  lattice,  with  lattice
spacing  of  a=1  fm.  Simulation  results  for  1000  events were
presented.  A  classification  parameter  (space-time  integrated
phases)  was  used  to distinguish between different events.
Evolution of spatially integrated phases,   different
modes, indicate that , on the average,
initial $\Theta$=0 state
do not evolve into a finite $\Theta$ state. Even in the most favored event,
a non-trivial $\Theta$ state is not produced. The phase remain uncorrelated
at later times also.
  However, as the sample space is not large,
and as the $\Theta$ vacua states are rare events, it is not possible to
conclude definitely that $\Theta$ vacua states will not be formed in heavy ion collisions. 
Situation is different if initially
 $\Theta$ is nonzero. In this case, initial  random  phases  quickly
evolve  to  values  dictated  by  the  potential and continue to oscillate around it.
Correlation studies also show  that  the
initial  random phases quickly develop large correlation, forming
a large parity odd bubble. Fourier analysis of the modes indicate
that initial $\Theta \neq 0$ states survive the evolution as a  coherent  superposition  of  a
number of modes.

\begin{figure}
\centerline{\psfig{figure=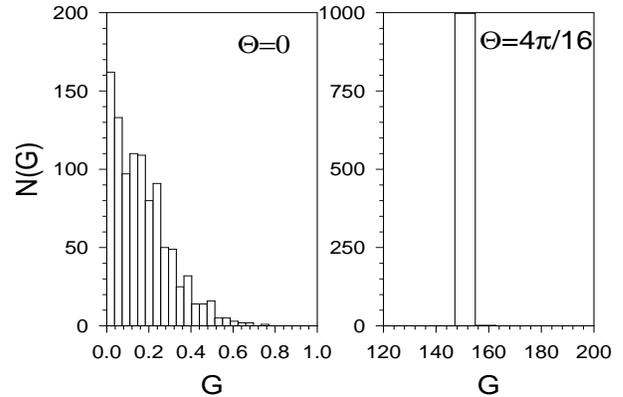,height=10cm,width=8cm}}
\vspace{-3 cm}
\caption{Distribution  of the classification parameter G, for 
$\Theta$=0 and  $\Theta$=4$\pi$/16.}
\end{figure}

\begin{figure}
\centerline{\psfig{figure=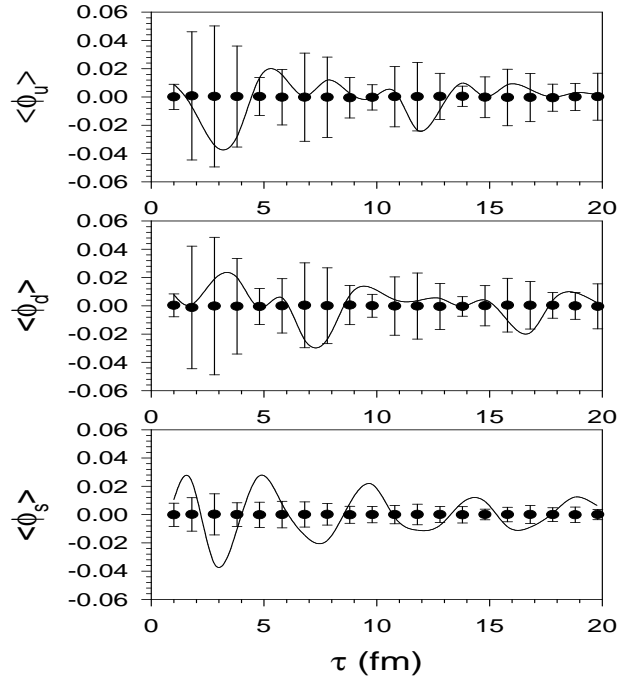,height=10cm,width=10cm}}
\caption{
Evolution  of  the  ensemble averaged, spatially  averaged  chiral  phases,
$\phi_u$, $\phi_d$ and $\phi_s$ for initial $\Theta$=0 (black
circles
with  error  bars). 
The solid lines show the evolution of the spatially averaged phases in
 the most favored event.
}
\end{figure}

\begin{figure}
\centerline{\psfig{figure=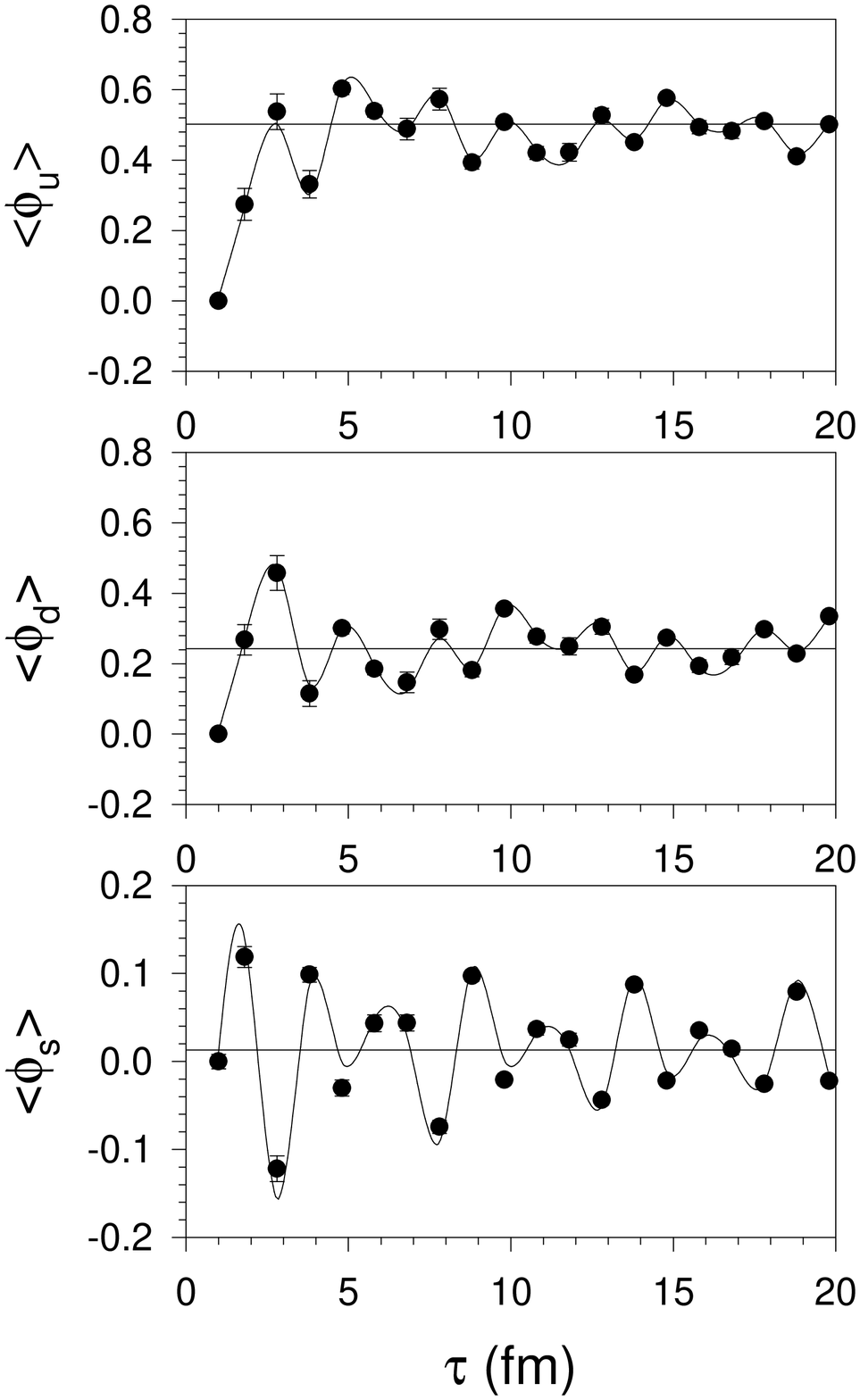,height=10cm,width=10cm}}
\caption{Same as fig.2 for $\Theta$=4$\pi$/16.}
\end{figure}

\begin{figure}
\centerline{\psfig{figure=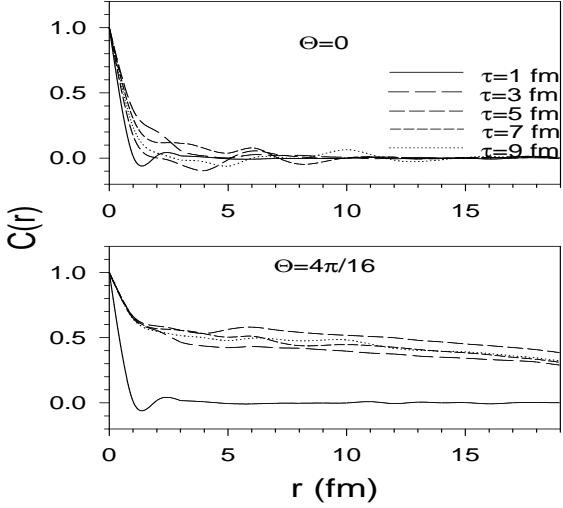,height=10cm,width=10cm}}
\vspace{-2.5cm}
\caption{Correlation   function   at   different  times  for  
$\Theta$=0 and  $\Theta$=4  $\pi$/16,  in  the  most  favored
event.}
\end{figure}

\begin{figure}
\centerline{\psfig{figure=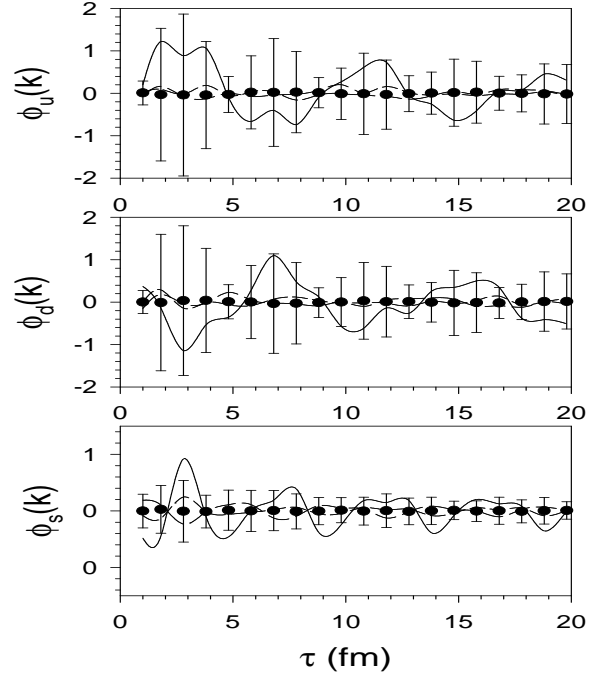,height=10cm,width=10cm}}
\caption{Evolution    of    the ensemble averaged  $\phi_i(k)$, k=6.2 MeV
 (black circles with error bar), for initial $\Theta =0$. The solid,
dashed, dash-dashed lines are the evolution of $\phi_i(k)$ 
in the most favored event, for
 k=6.2,  18.5  and  30.9
MeV respectively.}
\end{figure}

\begin{figure}
\centerline{\psfig{figure=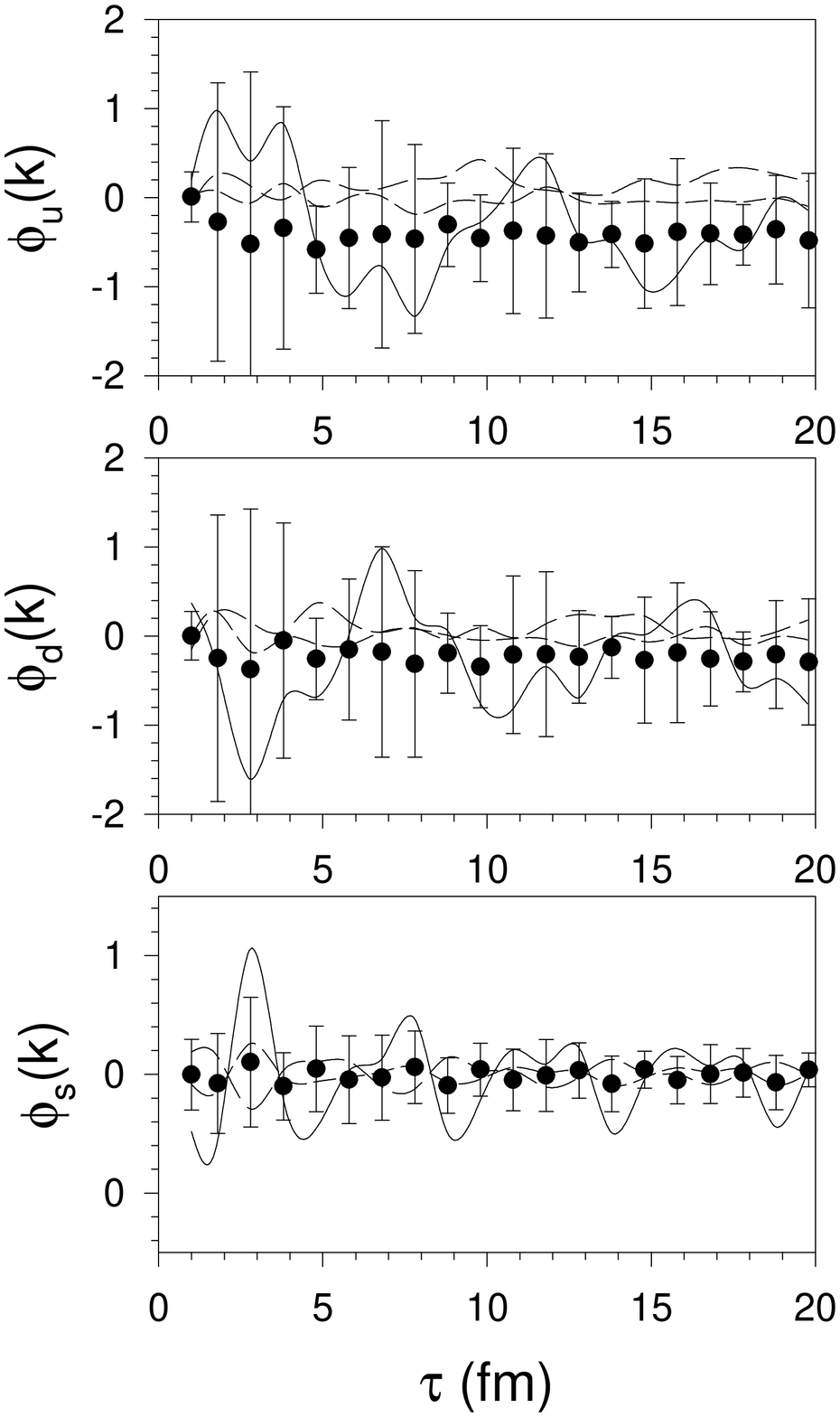,height=10cm,width=10cm}}
\caption{Same as Fig. 5 for  $\Theta$=4 $\pi$/16}
\end{figure}
\end{document}